\documentclass[a4paper,11pt]{article}
\usepackage{pos}

\title{The muon $g-2$ with four flavors of staggered quarks}
\ShortTitle{The muon $g-2$ with four flavors of staggered quarks}

\author*[a]{C.\ Aubin}
\author[b]{T.\ Blum}
\author[c,d]{M.\ Golterman}
\author[d]{S.\ Peris}

\affiliation[a]{Dept.\ of Physics \&
Engineering Physics, Fordham Univ.,\\ 
Bronx, NY 10458, USA}

\affiliation[b]{Dept.\ of Physics, Univ. of Connecticut,\\
Storrs, CT 06269, USA}

\affiliation[c]{Dept.\ of Physics and Astronomy,\\
San Francisco State Univ., San Francisco, CA 94132, USA}

\affiliation[d]{Dept.\ of Physics and IFAE-BIST, Univ. Aut\`onoma de Barcelona,\\
 E-08193 Bellaterra, Barcelona, Spain}

\emailAdd{caubin@fordham.edu}
\emailAdd{tblum@phys.uconn.edu}
\emailAdd{maarten@sfsu.edu}
\emailAdd{peris@ifae.es}

\abstract{We present updated results for the light-quark 
connected part of the leading hadronic contribution to the 
muon $g-2$ from configurations with 2+1+1 flavors of HISQ 
quarks using the time-momentum representation of the 
electromagnetic current correlator. We have added statistics 
on two ensembles as well as a fourth lattice spacing 
using configurations that have been generated by the MILC 
collaboration at the physical pion mass. Additionally we account 
for the leading finite-volume and taste-breaking effects 
using Staggered Chiral Perturbation Theory at NNLO.}

\FullConference{%
 The 38th International Symposium on Lattice Field Theory, LATTICE2021
  26th-30th July, 2021
  Zoom/Gather@Massachusetts Institute of Technology
}

\begin{document}
\maketitle

\section{Introduction}\label{sec:intro}

The muon anomalous magnetic
moment $a_\mu = (g-2)/2$ has become a leading 
candidate quantity with which to test the Standard
Model. The latest experimental result from Fermilab
this past April \cite{fermilabmuon}, when combined with
the previous result from Brookhaven \cite{Bennett:2006fi}, 
now shows
a 4.2$\sigma$ deviation 
from the Standard Model prediction (for a summary of the results
from the $g-2$ theory initiative, both 
from phenomenology and the lattice, see Ref.~\cite{AOYAMA20201}). 
The BMW collaboration recently found a smaller 
discrepancy with experiment, $1.5\sigma$ \cite{Borsanyi_2021}, when 
their sub-percent lattice QCD 
value for the HVP contribution is used 
instead of the data-driven value from the Muon g-2 Theory 
Initiative~\cite{AOYAMA20201}.
The uncertainty in the Fermilab result will only 
decrease in the next couple of years, and it is important
for the Standard Model calculation, which until recently has 
primarily come from the dispersive estimate,
to keep pace. 

It has become increasingly clear that a first-principles
approach to determining the QCD contributions using
lattice calculations is possible, given 
improvements in 
simulations during the last ten to fifteen years.
Improving statistics on the
lattice data is crucial of course, 
but also correcting for the many systematics is
necessary to produce a reliable result. 

We provide an update on our previous 
results \cite{Aubin:2019usy,uslat2019}
for the leading hadronic contribution to the 
muon $g-2$ (coming from the light quarks only), 
with a continuing focus on the
systematics that arise when using staggered
quarks in a finite volume. 
To do so, we have calculated the 
finite-volume and taste-breaking 
effects to next-to-next-to leading order (NNLO) 
in staggered chiral perturbation theory (ChPT).

\section{Simulation details}\label{sec:details}

We obtain the leading hadronic contribution to the muon
$g-2$ from the expression
\begin{equation}\label{eq:amu}
    a_\mu^{\rm HVP} 
    = 
    4 \alpha^2
    \int_0^\infty dq^2 f(q^2) \hat\Pi(q^2),
\end{equation}
where $f(q^2)$ is the kernel defined in Ref.~\cite{Blum:2002ii},
and $\hat \Pi(q^2) \equiv \Pi(0) - \Pi(q^2)$ is the
subtracted hadronic vacuum polarization, 
coming from the Fourier transform of the vector
two-point function (we use the conserved vector
current in our calculation). 
As is common with current lattice calcuations, we
use the time-momentum representation \cite{Bernecker:2011gh},
\begin{eqnarray}\label{eq:corr}
    \Pi(0) -\Pi(q^2) = \sum_t \left(
    \frac{\cos{qt} -1}{q^2} +\frac{1}{2} t^2
    \right) C(t),\quad 
    C(t) = \frac{1}{3}\sum_{\vec x,i}\langle 
    j^i(\vec x, t)j^i(0)\rangle,
\end{eqnarray}
where $C(t)$ is the Euclidean time correlation function, 
averaged over spatial directions. Equation~(\ref{eq:amu}) becomes
$a_{\mu}^{\rm HVP}(T) = 2\sum_{t=0}^{T/2} w(t) C(t)$,
with the weight
\begin{equation}\label{eq:kernel}
    w(t) = 4 \alpha^2 \int_{0}^{\infty} {d \omega^2}
    f(\omega^2)\left[\frac{\cos{\omega t} -1}{\omega^2} 
    +  \frac{t^2}{2}\right].
\end{equation}
% The weight is sometimes modified by replacing 
% the continuum Euclidean momentum-squared with 
% its lattice version $\hat w(t)$, where the $\omega^2$ in the 
% denominator of the first term in square brackets 
% is replaced with $[2\sin(\omega/2)]^2$ \cite{Blum:2018mom}. 

To reduce noise in our results, we use a combination
of full volume low-mode averaging and all-mode averaging,
to calculate the correlator in Eq.~(\ref{eq:corr}) 
\cite{Giusti_2004,DEGRAND2004185,Blum:2012uh,Blum:2018mom}. While 
we do not include the details of this implementation here, 
they are discussed in Ref.~\cite{Aubin:2019usy}. 

\begin{table}[bt]
\begin{center}
\begin{tabular}{cccccccc}
$m_\pi\ ({\rm MeV})$ &
$a\ ({\rm fm})$ & $L^3$ & $L\ ({\rm fm})$
& $m_\pi L$ &  LM & \#\ confs & traj. sep.\\
\hline
133 & 0.15148(80) & $32^3$ & 4.83 & 3.26 & 8000 & 48 & 40\\
130 & 0.08787(46) & $64^3$ & 5.62 & 3.66 & 8000 & 38 & 100\\
134 & 0.05684(30) & $96^3$ & 5.46 & 3.73 & 8000 & 35 & 60\\
\hline
\end{tabular}
\caption{Lattice simulation details for the additional
results for our calculation.}\label{tab:newdetails}
\end{center}
\end{table}

% \begin{table}[bt]
% \begin{tabular}{cccccccc}
% $m_\pi\ ({\rm MeV})$ &
% $a\ ({\rm fm})$ & $L^3$ & $L\ ({\rm fm})$
% & $m_\pi L$ &  LM & \#\ confs & traj. sep.\\
% \hline
% 133 & 0.12121(64) & $48^3$ & 5.82 & 3.91 & 4000 & 48 & 40\\
% 130 & 0.08787(46) & $64^3$ & 5.62 & 3.66 & 4000 & 38 & 12\\
% 134 & 0.05684(30) & $96^3$ & 5.46 & 3.73 & 6000 & 35 & 48\\
% \hline
% \end{tabular}
% \caption{Lattice simulation details for the old
% results for our calculation.}\label{tab:newdetails}
% \end{table}

We use  configurations generated by the MILC Collaboration 
\cite{Bazavov:2014wgs}
with 2+1+1 flavors of HISQ quarks. The new data comes from 
the configurations shown in 
Table~\ref{tab:newdetails}, which we compare to our older results
(not shown here, but they are shown in
Ref.~\cite{Aubin:2019usy}). We have new computations on the
two finest lattice spacings ($a\approx 0.06$ and 0.09 fm), and have
added the coarser $a\approx0.15$~fm lattice spacing, all 
near the physical pion mass. 
All but the coarsest lattice spacing have a 
spatial volume of around $(5.5~\rm{fm})^3$ (the $a=0.15$ fm 
ensemble has $L^3\sim (4.83~{\rm fm})^3$), and 
$m_\pi L\approx 3.3 - 3.7$. 
For all of the newer results, we use 8000 low modes (as opposed
to either 4000 or 6000 low modes in our original results), and
we separate our lattices by 40-100 trajectories as opposed
to 12-48 in our original results.

\section{Finite-volume chiral perturbation theory}\label{sec:chpt}

In order to study the leading staggered taste-breaking
effects we calculate
the vector correlator to two loops 
(NNLO) in staggered ChPT \cite{Aubin:2003mg,us2020}. 
In our earlier result \cite{Aubin:2019usy,uslat2019}, we 
used a hybrid approach: we calculated
the one-loop corrections in staggered ChPT, extrapolated to the 
continuum, then applied the two-loop continuum finite volume 
corrections. Here however, we perform 
the full staggered ChPT calculation (in Euclidean space) 
at two loops as was done by the BMW 
collaboration \cite{Borsanyi_2021}. 

After a straightforward (albeit lengthy) calculation,
we obtain at NNLO,
\begin{eqnarray}\label{eq:coft}
    C(t) & = &  
    \frac{1}{48 L^3}\sum_{\vec p}\sum_X\frac{{\vec p}^2}{E^2_X(p)}\, e^{-2E_X(p)t} \Biggl(1
-\frac{1}{4f^2}\sum_Y D_Y(0) - \frac{16\ell_6(\vec p^2+m_X^2)}{f^2}
\nonumber
\\
&&
+\frac{1}{24f^2}\,\frac{1}{L^3}\sum_{\vec q}\sum_Y\frac{{\vec q}^2}{E_Y(q)}
\,\frac{1}{\vec q^2-\vec p^2+m_Y^2-m_X^2}\Biggr),
\end{eqnarray}
where the sums over $X$ and $Y$ run over the 
16 pion tastes for staggered quarks, and 
\begin{equation}\label{eq:Dofzero}
    D_Y(0) = 
    \frac{1}{L^3}\sum_{\vec k } \frac{1}{2E_Y(k)}, \qquad
    E_Y(k) = \sqrt{k^2 + m_Y^2}.
\end{equation}
The sums over $\vec p$ and $\vec k$ are over 
the momenta $2\pi{\vec n}/L$ with ${\vec n}$ a three-vector
of integers in a box with periodic boundary conditions. 
We define the renormalized $\ell_6^r$ by
\begin{equation}\label{l6ren}
    \ell_6 = \ell_6^r(\mu)
    -\frac{1}{3}\,\frac{1}{16\pi^2}
    \left(\frac{1}{\epsilon}-\log\mu - 
    \frac{1}{2}(\log{(4\pi)}-\gamma+1)\right)\ .
\end{equation}

In order to extract the
finite-volume difference between
the infinite-volume result for $a_\mu^{\rm HVP}$ and the 
finite-volume result,
$\Delta a_\mu^{\rm HVP} =
\left[\lim_{L\to\infty} a_\mu^{\rm HVP}(L)\right]
- a_\mu^{\rm HVP}(L)$ in ChPT, we apply
the Poisson Summation formula to Eq.~(\ref{eq:coft}).
The resulting difference
$\Delta a_\mu^{\rm HVP}$ follows, and after some simplification
we perform the sums and integrals numerically.

With the parameters for our ensembles, we have
calculated $\Delta a_\mu^{\rm HVP}$ both for the full 
result, as well as for the intermediate window 
method~\cite{Blum:2018mom} where we have
\begin{eqnarray}
    a_\mu^{\rm HVP, win} & = & 
    2\sum_{t=0}^{T/2} C(t) w(t) (\Theta(t,t_0,\Delta)-\Theta(t,t_1,\Delta))
    ,\\
    \Theta(t,t',\Delta) &
    = &\frac{1}{2} (1+\tanh((t-t')/\Delta)) ,
\end{eqnarray}
with $t_0 =0.4$~fm, $t_1 = 1.0$~fm, and $\Delta= 0.15$~fm, 
as this allows us to compare 
with other lattice results. We 
show $\Delta a_\mu^{\rm HVP, win}\times10^{10}$ for the window method in 
Table~\ref{tab:FVcorrections} for 
each of our ensembles (only at NLO),
where we breakdown the finite-volume corrections due
to the taste-pseudoscalar pion mass, the additional corrections
that arise from staggered taste breaking, a correction needed
to account for retuning the pion mass, and the final column is
the total correction (from summing the previous three columns).
We have the full 
NNLO corrections for the entire momentum range, however
the numerical corrections for the window are more complicated
to evaluate, so these were not included in this update.

\begin{table}[t]
\begin{center}
\begin{tabular}{ccccccc}
&& & $m_{\pi_5}$\     &   \textrm{taste-breaking } &  $ m_\pi $  & \\
$a$ (fm) & {\rm Volume} & $m_{\pi_5} {\rm (MeV)} $&   {\rm FV\ corr.}    &  {\rm in\ FV}  &   {\rm   retuning     }   & {\rm  total}\\
\hline
0.05684 & $96^3$ & 134 &   0.727538 &     0.759853   &     -0.0687552   &  1.4186\\
0.08787 & $64^3$ & 130 &   0.697276 &     3.51669   &       -0.516077    &   3.6979\\
0.12121 & $48^3$ & 133 &  0.560572  &    7.99304    &      -0.21689      &   8.3367\\
0.15148 & $32^3$ & 133 &  1.24171    &  10.3470      &      -0.186605    & 11.4021\\
\hline
\end{tabular}
\caption{The various corrections $\Delta a_\mu^{\rm HVP, win}\times10^{10}$ 
% coming from finite-volume effects 
for the intermediate window at NLO from various sources: 
finite-volume effects due to the taste-pseudoscalar
pion mass, additional corrections from staggered
taste breaking, a 
correction need to retune the pion mass, 
and the total corrections (respectively). 
}\label{tab:FVcorrections}
\end{center}

\end{table}

\section{Results \& Conclusions}\label{sec:conc}

\begin{figure}[b]
    \includegraphics[width=\textwidth]{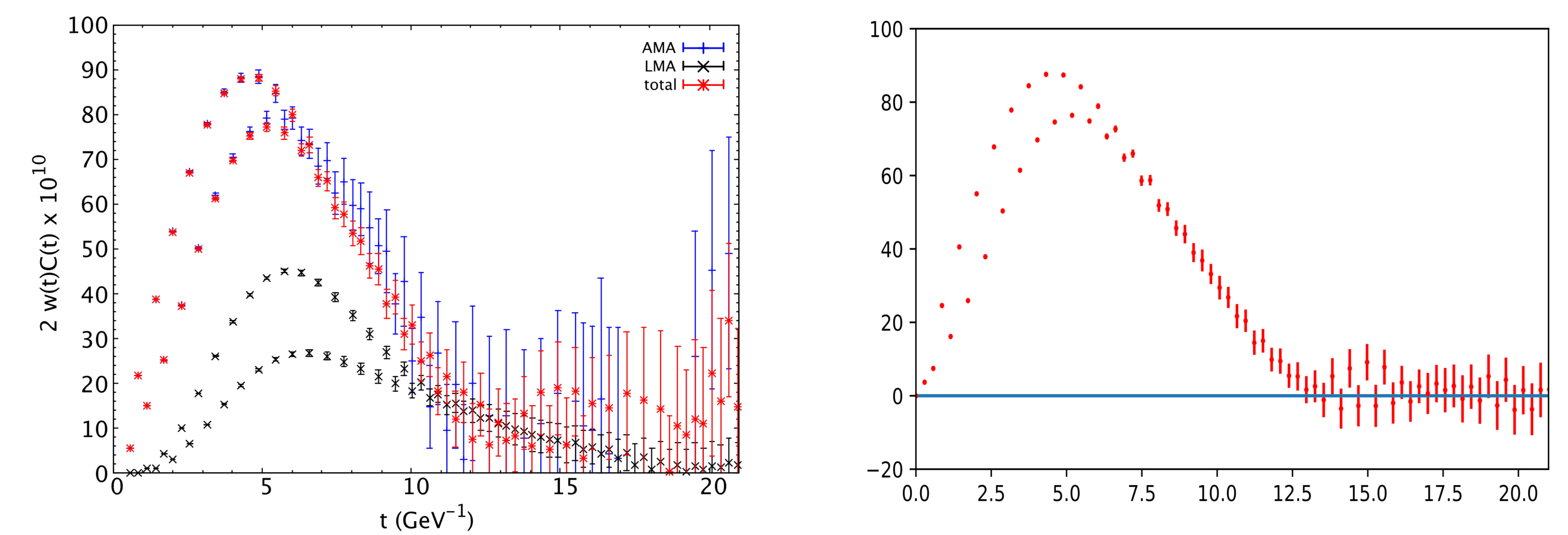}
    \caption{A comparison of the summand for our original 
    data vs.\ new data for the $96^3$ ensemble.}\label{fig:compare}
\end{figure}

For the $96^3$ ensemble, we compare our 
older results for the summand to the newer
results in Fig.~\ref{fig:compare}. On the left is the old
data with 23 configurations and 4000 low modes
(in red is the total result, combining the LMA and AMA 
results), and on the right is the new data with 35 
configurations and 8000 low modes. One can see
the marked reduction in error bars with the new data. 

To extract $a_\mu^{\rm HVP}$ from our data more precisely, we use
the bounding method of 
Refs.~\cite{Blum:2018mom,Borsanyi:2017zdw}. We set
bounds on the correlator for when $t$ is greater
than a time $T$: $0 < C(t) < C(T) e^{-E_0(t-T)}$, 
where %the lowest energy state in the vector channel is 
$E_0=\infty$ (lower bound) or $2\sqrt{m_\pi^2+(2\pi/L)^2}$
(upper bound). 
At sufficiently large $T$ the bounds overlap, 
and an estimate 
for $a_\mu$ can be made which is
more precise than simply summing 
over the noisy long-distance tail. 

\begin{figure}[tb]
    \includegraphics[width=0.5\textwidth]{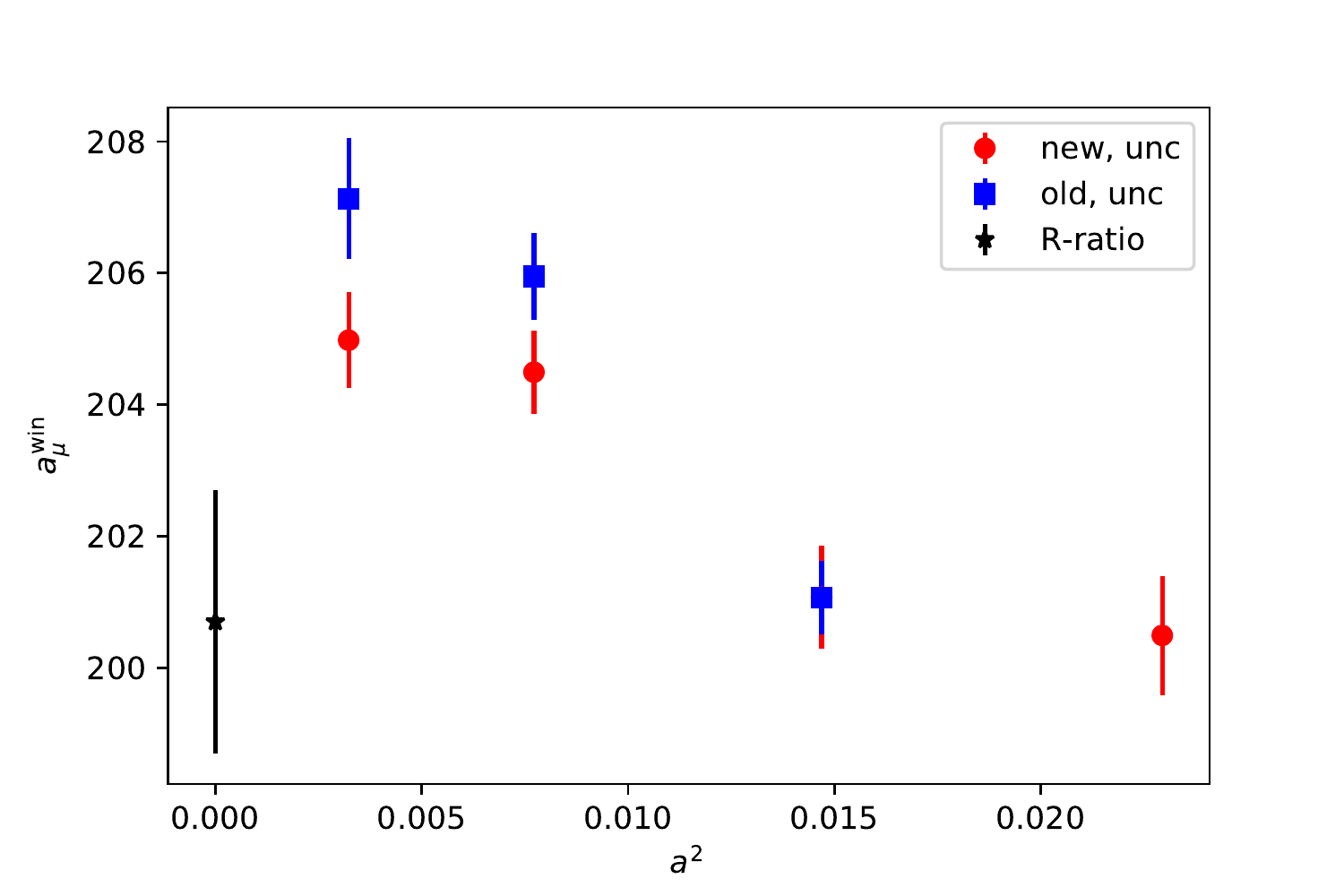}
    \hspace{0.03\textwidth}
    \includegraphics[width=0.5\textwidth]{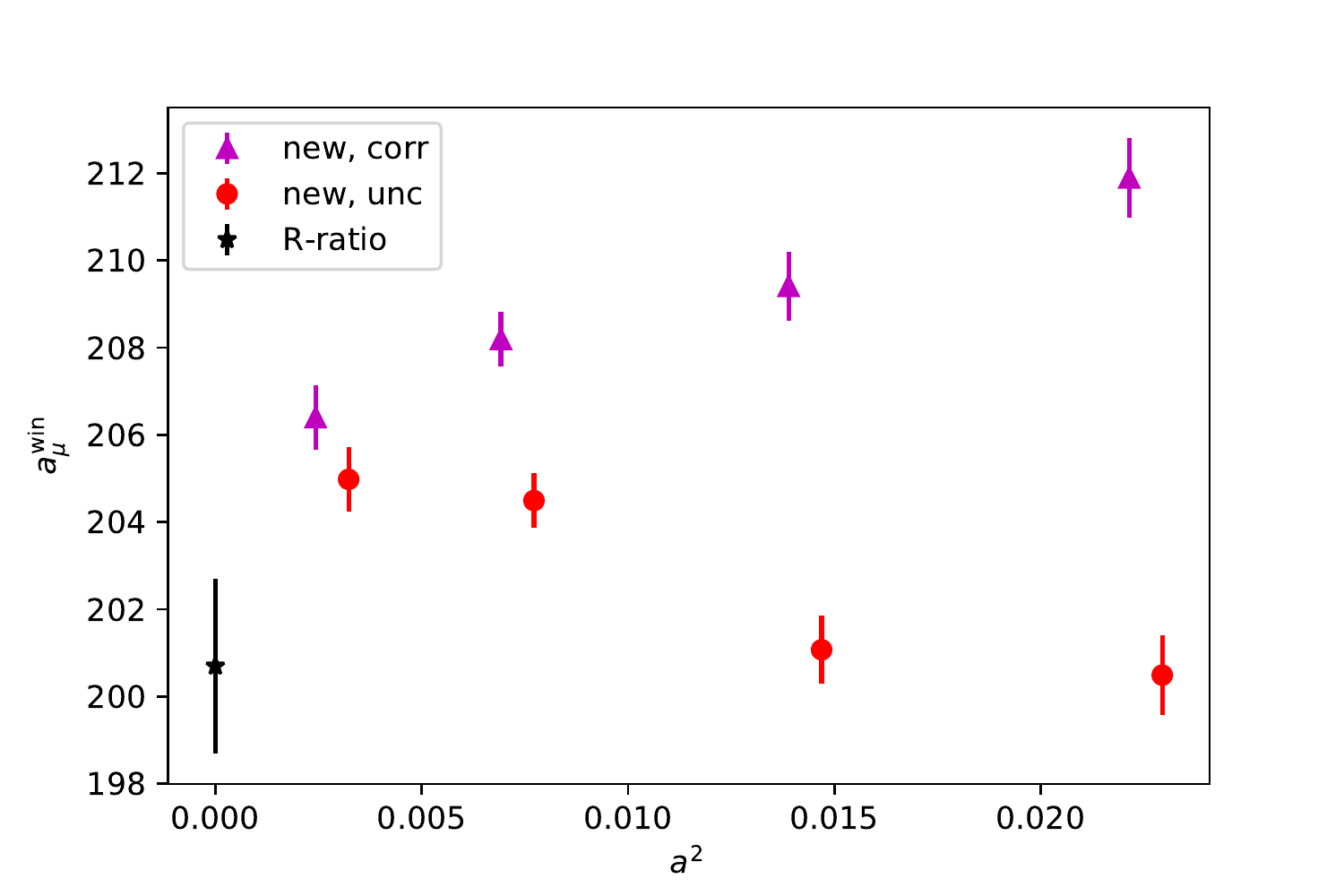}

    \centerline{(a)\hspace{0.5\textwidth}(b)}

    \caption{(a) A comparison of our old (circles) and new (squares)
    uncorrected results. (b) 
    A comparison of the new uncorrected (circles) and new corrected 
    (triangles) results.}\label{fig:alldata}
\end{figure}

We show in Fig.~\ref{fig:alldata} the extracted results 
for $a_\mu^{\rm HVP}$ using the bounding method. In
Fig.~\ref{fig:alldata}(a) we directly compare the 
old and new data, while in Fig.~\ref{fig:alldata}(b) we 
compare the uncorrected and corrected new results. 
Note, the $a = 0.12$~fm results are not new results, but are 
included in the fits, so we include that data point 
in the figure. The scaling behavior of the uncorrected
data is quite different than that of the corrected data:
the overall slope as $a^2 \to 0$ changes sign between the 
two, and the uncorrected data exhibits a kink 
between large and small lattice spacings.

\begin{figure}[tb]
    \includegraphics[width=0.49\textwidth]{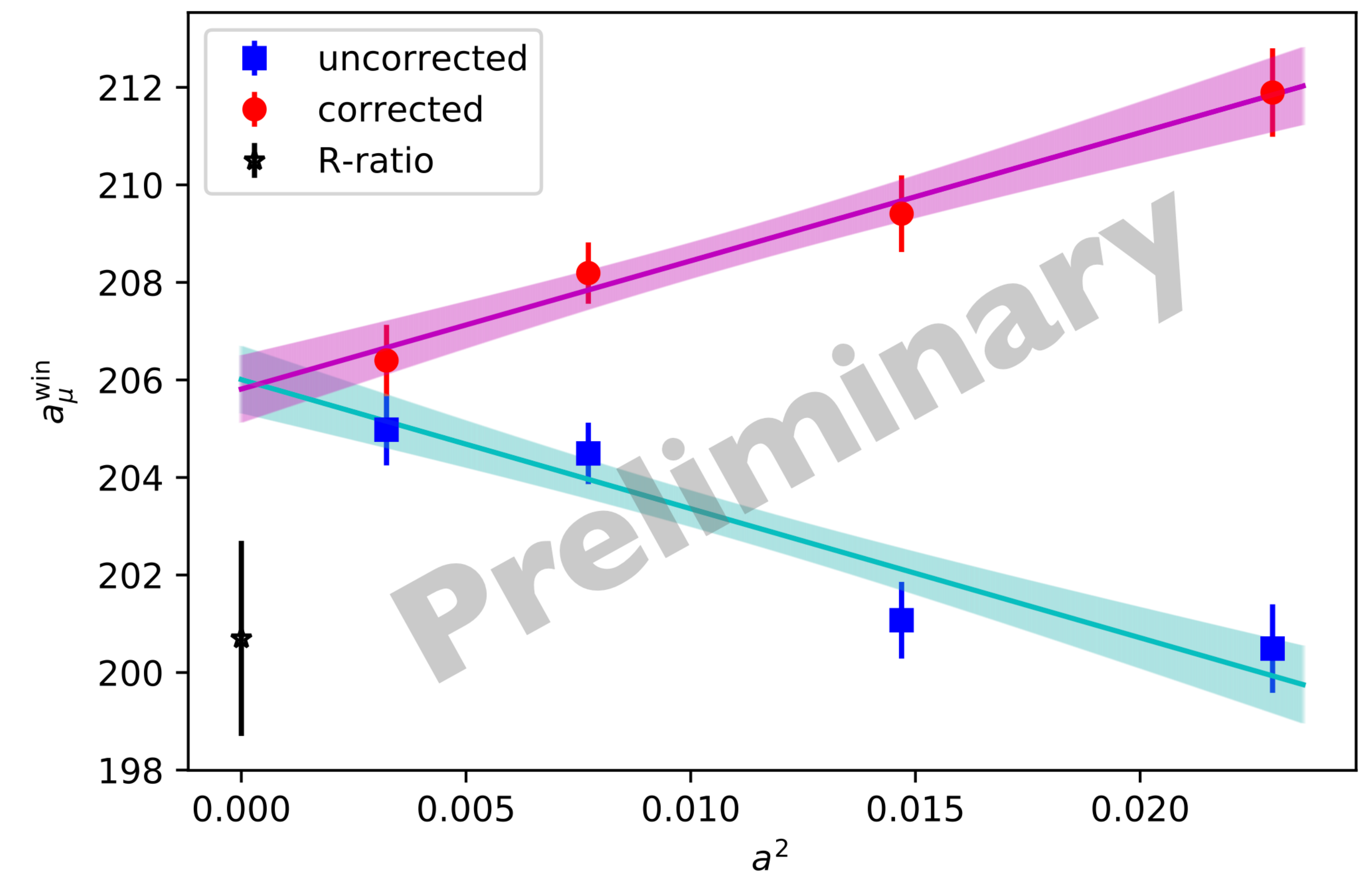}
    %\hspace{0.005\textwidth}
    \includegraphics[width=0.49\textwidth]{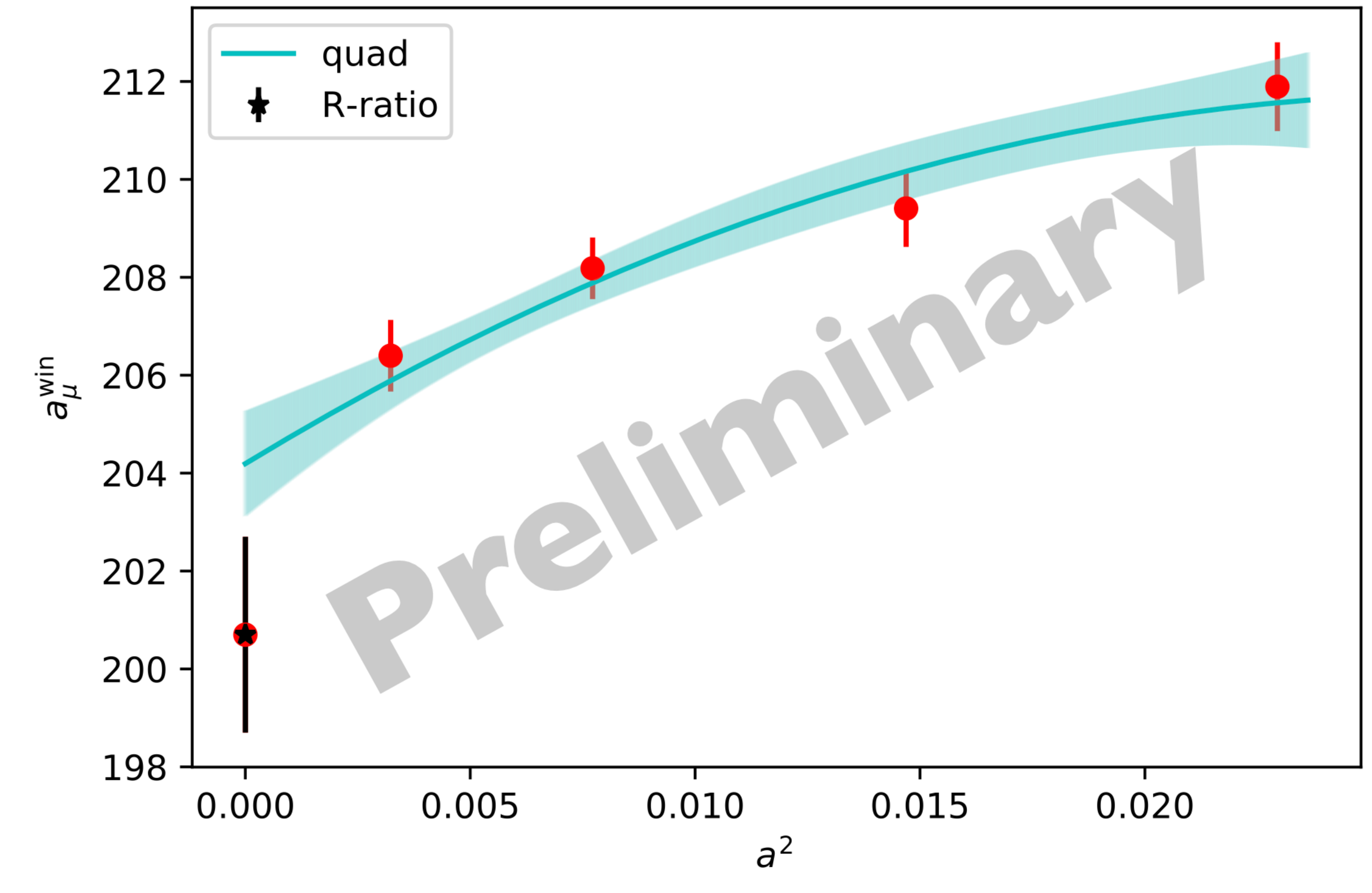}

        \centerline{(a)\hspace{0.5\textwidth}(b)}

\vspace{.1cm}

%\end{figure}
%\begin{figure}[tb]
    \includegraphics[width=0.49\textwidth]{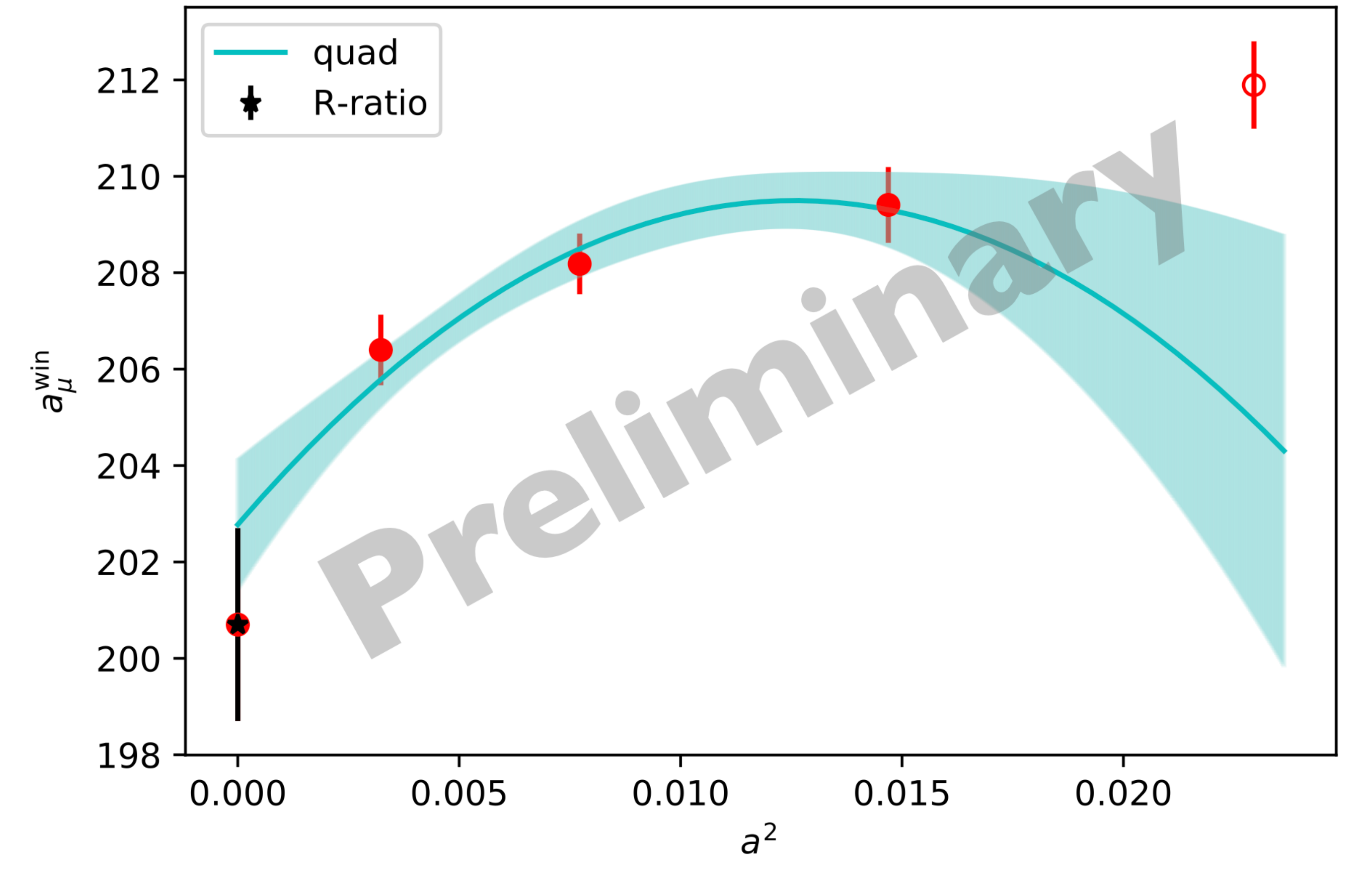}
    \includegraphics[width=0.49\textwidth]{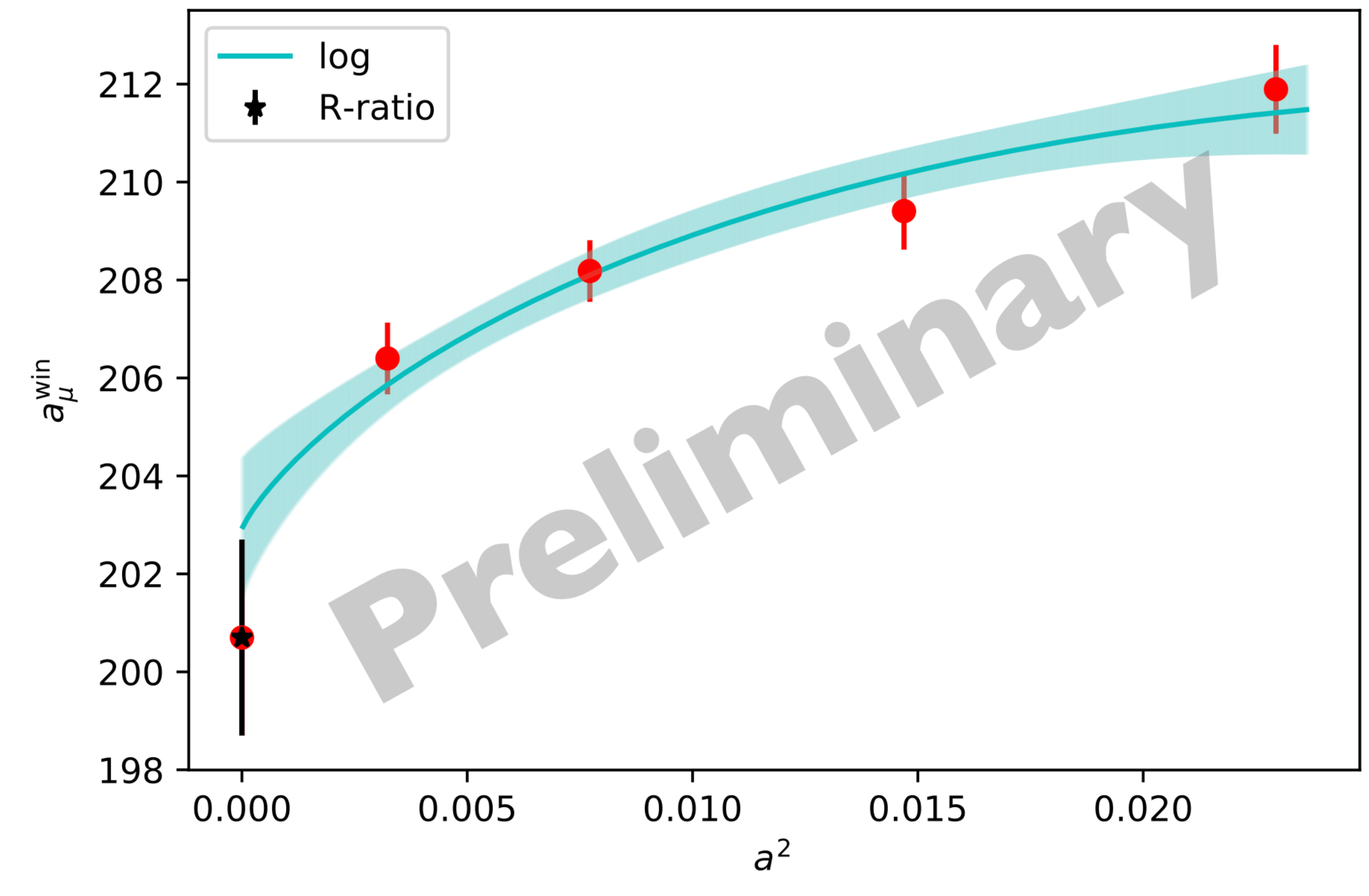}

    \centerline{(c)\hspace{0.5\textwidth}(d)}

    \caption{Various fits to the updated data. (a) compares
    a linear fit to the uncorrected and corrected data, while
    the other fits are only to corrected data, with (b) a 
    quadratic fit to all lattice points but not the $R$-ratio, 
    (c) a quadratic fit 
    to all but the coarsest lattice point + the $R$-ratio
    point, and (d) a perturbation-theory-inspired fit
    to all points (including the $R$-ratio).
    }\label{fig:fits}
\end{figure}

In Fig.~\ref{fig:fits}, we show various fits to the data. 
First we compare the original, uncorrected data to the corrected
data in Fig.~\ref{fig:fits}(a) with a simple linear 
fit in $a^2$.
The extrapolated results for both fits are quite consistent, but the
$\chi^2/$d.o.f.\
for the fit to the corrected data is better. In either case, the
extrapolated result is not consistent with the $R$-ratio (for this
value we use the result in our earlier work \cite{Aubin:2019usy}).
To check if our results could be consistent with the $R$-ratio
determination, we perform several fits with that as a data point.
Figs.~\ref{fig:fits}(b) \& (c) are quadratic fits in $a^2$, 
with and without the coarsest lattice spacing ($a\approx 0.15$~fm),
while Fig.~\ref{fig:fits}(d) is a fit inspired by perturbation
theory \cite{rainerPT}, with a 
logarithm term: $A + Ba^2 + C a^2 \log a^2$. Each of these 
fits have a reasonable $\chi^2/$d.o.f., and may indicate that 
the lattice data is not inconsistent with the $R$-ratio.

Our preliminary results suggest that simulations 
at even finer lattice spacings may be needed to 
determine whether the $R$-ratio and the lattice 
value are consistent.
 Additionally, finer lattice spacings
can help understand the cutoff effects (the odd behavior of the 
uncorrected data is evidence for this). 
Currently we are running 
on a larger $a\approx0.15$~fm ensemble (with $L=48$\footnote{Thanks 
to Andre Walker-Loud and 
the CalLat collaboration
for sharing these lattices.} instead of
$L=32$) to help 
get a better handle
on the finite volume effects.

\section*{Acknowledgments}

This work used the Extreme Science and Engineering 
Discovery Environment (XSEDE), which is supported by 
National Science Foundation grant number ACI-1548562.
at T.B.'s and M.G.'s work is supported by the U.S.\ 
Department of Energy, Office of Science, 
Office of High Energy Physics, under Grants No. DE-SC0010339 
and No. DE- SC0013682, respectively. S. P.\ 
has received funding from the Spanish Ministry 
of Science and Innovation 
(PID2020-112965GB-I00/AEI/ 10.13039/501100011033).

\bibliographystyle{JHEP}   % if natbib is available
\bibliography{refs2}

\providecommand{\href}[2]{#2}\begingroup\raggedright\begin{thebibliography}{10}

\bibitem{fermilabmuon}
{\bf Muon $g\ensuremath{-}2$} Collaboration, B.~Abi {\em et.~al.}, {\it
  Measurement of the positive muon anomalous magnetic moment to 0.46 ppm},
  {\em Phys. Rev. Lett.} {\bf 126} (Apr, 2021) 141801.

\bibitem{Bennett:2006fi}
{\bf Muon $g\ensuremath{-}2$} Collaboration, G.~Bennett {\em et.~al.}, {\it
  {Final Report of the Muon E821 Anomalous Magnetic Moment Measurement at
  BNL}},  {\em Phys.Rev.} {\bf D73} (2006) 072003.

\bibitem{AOYAMA20201}
T.~Aoyama {\em et.~al.}, {\it The anomalous magnetic moment of the muon in the
  standard model},  {\em Physics Reports} {\bf 887} (2020) 1--166.

\bibitem{Borsanyi_2021}
S.~Borsanyi, Z.~Fodor, J.~N. Guenther, C.~Hoelbling, S.~D. Katz, L.~Lellouch,
  T.~Lippert, K.~Miura, L.~Parato, K.~K. Szabo, and et~al., {\it Leading
  hadronic contribution to the muon magnetic moment from lattice qcd},  {\em
  Nature} {\bf 593} (Apr, 2021) 51–55,
  [\href{http://xxx.lanl.gov/abs/2002.1234}{{\tt arXiv:2002.1234}}].

\bibitem{Aubin:2019usy}
C.~Aubin, T.~Blum, C.~Tu, M.~Golterman, C.~Jung, and S.~Peris, {\it {Light
  quark vacuum polarization at the physical point and contribution to the muon
  $g-2$}},  {\em Phys. Rev. D} {\bf 101} (2020), no.~1 014503,
  [\href{http://xxx.lanl.gov/abs/1905.0930}{{\tt arXiv:1905.0930}}].

\bibitem{uslat2019}
C.~Aubin, T.~Blum, M.~Golterman, C.~Jung, S.~Peris, and C.~Tu, {\it {Hadronic
  vacuum polarization in finite volume using NNLO ChPT}},  {\em PoS} {\bf
  LATTICE2019} (2019) 102, [\href{http://xxx.lanl.gov/abs/1910.0509}{{\tt
  arXiv:1910.0509}}].

\bibitem{Blum:2002ii}
T.~Blum, {\it {Lattice calculation of the lowest order hadronic contribution to
  the muon anomalous magnetic moment}},  {\em Phys.Rev.Lett.} {\bf 91} (2003)
  052001.

\bibitem{Bernecker:2011gh}
D.~Bernecker and H.~B. Meyer, {\it {Vector Correlators in Lattice QCD: Methods
  and applications}},  {\em Eur.Phys.J.} {\bf A47} (2011) 148.

\bibitem{Giusti_2004}
L.~Giusti, P.~Hernandez, M.~Laine, P.~Weisz, and H.~Wittig, {\it Low-energy
  couplings of {QCD} from current correlators near the chiral limit},  {\em
  Journal of High Energy Physics} {\bf 2004} (apr, 2004) 013--013.

\bibitem{DEGRAND2004185}
T.~DeGrand and S.~Schaefer, {\it Improving meson two-point functions in lattice
  qcd},  {\em Computer Physics Communications} {\bf 159} (2004), no.~3
  185--191.

\bibitem{Blum:2012uh}
T.~Blum, T.~Izubuchi, and E.~Shintani, {\it {New class of variance-reduction
  techniques using lattice symmetries}},  {\em Phys.Rev.} {\bf D88} (2013)
  094503.

\bibitem{Blum:2018mom}
{\bf RBC, UKQCD} Collaboration, T.~Blum {\em et.~al.}, {\it {Calculation of the
  hadronic vacuum polarization contribution to the muon anomalous magnetic
  moment}},  {\em Phys. Rev. Lett.} {\bf 121} (2018), no.~2 022003.

\bibitem{Bazavov:2014wgs}
{\bf Fermilab Lattice, MILC} Collaboration, A.~Bazavov {\em et.~al.}, {\it
  {Charmed and light pseudoscalar meson decay constants from four-flavor
  lattice QCD with physical light quarks}},  {\em Phys. Rev.} {\bf D90} (2014),
  no.~7 074509.

\bibitem{Aubin:2003mg}
C.~Aubin and C.~Bernard, {\it {Pion and kaon masses in staggered chiral
  perturbation theory}},  {\em Phys. Rev.} {\bf D68} (2003) 034014.

\bibitem{us2020}
C.~Aubin, T.~Blum, M.~Golterman, and S.~Peris, {\it {Application of effective
  field theory to finite-volume effects in $a_\mu^{HVP}$}},  {\em Phys. Rev. D}
  {\bf 102} (2020), no.~9 094511,
  [\href{http://xxx.lanl.gov/abs/2008.0380}{{\tt arXiv:2008.0380}}].

\bibitem{Borsanyi:2017zdw}
{\bf BMW} Collaboration, S.~Borsanyi {\em et.~al.}, {\it {Hadronic vacuum
  polarization contribution to the anomalous magnetic moments of leptons from
  first principles}},  {\em Phys. Rev. Lett.} {\bf 121} (2018), no.~2 022002.

\bibitem{rainerPT}
N.~Husung, P.~Marquard, and R.~Sommer, {\it {Asymptotic behavior of cutoff
  effects in Yang\textendash{}Mills theory and in Wilson\textquoteright{}s
  lattice QCD}},  {\em Eur. Phys. J. C} {\bf 80} (2020), no.~3 200,
  [\href{http://xxx.lanl.gov/abs/1912.0849}{{\tt arXiv:1912.0849}}].

\end{thebibliography}\endgroup

\end{document}